# Optimizing the sensitivity of high repetition rate broadband transient optical spectroscopy with modified shot-to-shot detection

Siedah J. Hall,[1,2] Peter J. Budden,[1,2] Anne Zats,[1] Matthew Y. Sfeir[1,2,a]

[1]*Photonics Initiative, Advanced Science Research Center, City University of New York, New York, New York 10031, USA*
[2]*Department of Physics, The Graduate Center, City University of New York, New York, New York 10016, USA*

[a]Author to whom correspondence should be addressed: msfeir@gc.cuny.edu.

A major limitation of transient optical spectroscopy is that relatively high laser fluences are required to enable broadband, multichannel detection with acceptable signal-to-noise levels. Under typical experimental conditions, many condensed phase and nanoscale materials exhibit fluence-dependent dynamics, including higher-order effects such as carrier-carrier annihilation. With the proliferation of commercial laser systems, offering both high repetition rates and high pulse energies, has come new opportunities for high sensitivity pump-probe measurements at low pump fluences. However, experimental considerations needed to fully leverage the statistical advantage of these laser systems has not been fully described. Here we demonstrate a high repetition rate, broadband transient spectrometer capable of multichannel shot-to-shot detection at 90 kHz. Importantly, we find that several high-speed cameras exhibit a time-domain fixed pattern noise resulting from interleaved analog-to-digital converters that is particularly detrimental to the conventional "ON/OFF" modulation scheme used in pump-probe spectroscopy. Using a modified modulation and data processing scheme, we achieve a noise level of $10^{-5}$ OD in four seconds, an order of magnitude lower than for commercial 1 kHz transient spectrometers for the same acquisition time. We leverage the high sensitivity of this system to measure the differential transmission of monolayer graphene at low pump fluence. We show that signals on the order of $10^{-6}$ OD can be measured, enabling a new data acquisition regime for low-dimensional materials.

## I. INTRODUCTION

Transient optical spectroscopy based on the "pump-probe" technique is a ubiquitous tool for observing light driven dynamic processes in molecular and condensed matter systems. These measurements, which can be expressed in terms of a differential transmission ($\Delta T/T$), absorption ($\Delta A$), or reflection signal ($\Delta R/R$), represent a special case of four-wave mixing and generate a nonlinear signal proportional to the third order polarization. In its simplest implementation, two distinct laser pulses derived from the same master source are overlapped in space and time on a sample. A strong interaction with a "pump" pulse, typically in resonance with a fundamental excitation, drives the system out of equilibrium. An additional weak interaction with a degenerate or non-degenerate "probe" pulse occurs following an appropriate delay and interrogates time dependent changes in the refractive index of the material. For phase matching along probe direction, the resulting nonlinear signal can be expressed in a form analogous to the Beer-Lambert Law such that $\Delta T/T \approx -\Delta \alpha L = -\ln(10)\Delta A$, where $\alpha$ is the absorption coefficient and $L$ is the path length. Signatures of different excited state configurations, including their lifetimes and decay products, can be readily determined as a function of the probe wavelength, including spin conversion, exciton dissociation, optical gain, and charge transport processes.[1,2]

Over many years of development, a consensus has emerged that optimal implementations for transient optical spectroscopy incorporate supercontinuum probe pulses with octave spanning spectral content in the visible and near-infrared[3] in conjunction with shot-to-shot multiplexed detection. Detailed implementations of these approaches have been widely reported[4] and several commercial platforms exist that offer fully integrated transient optical spectrometers based on these principles.[5–7] There are numerous advantages of broadband multichannel over single-channel detection, including facilitating multidimensional data processing methods such as global and target analysis.[8,9] These methods permit the deconvolution of characteristic signals associated with distinct transient species, improve the precision of kinetic fits, and provide useful noise filtering capabilities. Furthermore, wavelength resolved detection facilitates the observation of subtle spectral shifts due to exciton correlation and migration dynamics.[10–12] In addition to improving spectral resolution, multiplexed detection enables high time resolution for highly chirped probe pulses without complex dispersion compensation schemes.[13] While many of these advantages are obtained with arbitrary data collection schemes, the ideal approach is to use shot-to-shot multichannel detection, in which data sets are derived from pairs of consecutive laser shots without averaging over multiple laser pulses.[14] This approach minimizes noise due to laser fluctuations and drift and achieves the highest sensitivity.[15,16]

Despite the great success of broadband transient optical measurements in advancing materials science, further improvements require large scale changes in the underlying technologies, including higher repetition rate lasers, faster multichannel detectors, and jitter-free high speed pulse modulation techniques. Historically, Ti:Sapphire based amplified laser systems have been used to build broadband ultrafast spectroscopy systems. These stable solid-state lasers support large pulse bandwidths, can readily provide multi-millijoule pulse energies, and are convenient for pumping optical parametric amplifiers (OPAs) and for supercontinuum generation. However, these systems are limited to few kHz repetition rates at room temperature, necessitating long acquisition times and high pump fluences to obtain sufficient signal-to-noise ratios. These measurement conditions are not ideal, as at typical fluence levels in conventional measurements ($> 1$ $\mu J/cm^2$), many solid-state and nanoscale materials exhibit carrier annihilation or other higher-order effects that obscure the intrinsic carrier dynamics.

The recent proliferation of commercial high repetition rate amplified laser systems based on Yb:KGW gain media has encouraged the development of a new generation of high-speed broadband transient optical measurements with enhanced sensitivity. The high pulse energies (0.1 – 1 mJ) of these laser systems facilitate wavelength tuning via optical parametric amplification as well as supercontinuum generation, while the high repetition rates (hundreds of kHz or higher) and better shot-to-shot stability can be harnessed to achieve improved signal-to-noise levels compared to Ti:Sapphire-based measurements over the same acquisition time.[17] Pioneering work demonstrated that shot-to-shot detection of supercontinuum pulses using array detectors could be achieved at 100 kHz line rates and used to build high speed transient optical measurement systems.[14] However, this approach is based on two pieces of equipment that are no longer commercially available: a continuous wave pumped low pulse energy Ti:Sapphire amplifier (Coherent RegA) and high speed CCD camera (e2v Aviiva EM4). Furthermore, this approach necessitated custom mechanical modulation with complex synchronization electronics. A generalized approach using modern hardware is highly desirable but presents several challenges. Mechanical modulation at 100 kHz is problematic as it requires extremely small apertures and results in noisy operation, large air currents, and high jitter. Furthermore, while there are tantalizing reports of signal-to-noise gains in multi-dimensional electronic[17] and time-resolved Raman measurements[18] as repetition rates are increased, the ultimate sensitivity of transient optical spectroscopy with high speed line cameras has not been quantified.[19]

Here we present an updated implementation of a transient optical spectrometer that achieves shot-to-shot detection at 90 kHz without mechanical modulation. This approach is generalizable, compact, and cost effective as it directly reads and processes the output from an inexpensive line camera with a USB interface, i.e., no specialized data acquisition hardware is needed. Importantly, we find that the implementation of analog-to-digital conversion (ADC) processes in many high-speed cameras is incompatible with conventional "on-off" modulation and data processing schemes for transient spectroscopy. Here, we demonstrate how time-domain fixed pattern noise spurs in high-speed line cameras can be minimized using a modified burst modulation approach. This modified modulation scheme allows us to achieve an order of magnitude reduction in noise compared to a commercial Ti:Sapphire based transient absorption spectrometer for the same acquisition time. We use this system to measure the carrier dynamics of monolayer graphene over a broad range of probe wavelengths for a pump fluence below 1 $\mu J/cm^2$ per pulse, representing a previously inaccessible data acquisition regime.

## II. DATA ACQUISITION AND PROCESSING

### A. Optical Layout

Our transient optical spectrometer is based on a commercial 40 W Yb:KGW amplified laser system (Light Conversion Carbide CB3), with a maximum pulse energy of 0.4 mJ at 100 kHz, center wavelength of 1030 nm, and pulse duration of approximately 230 fs. We reduced the repetition rate of the laser amplifier to 90 kHz to match the speed of our fast line cameras. To facilitate pulse picking of high energy pump pulses via an electro-optic modulator, half of the fundamental (20 W) is split off from the amplifier before recompression and passed through an external pulse picker. Following electro-optic modulation, an external grating compressor is used to recompress pulses to their transform limit (**Fig. 1**). We note that a simpler approach, in which a pulse picker is applied directly to compressed pulses, is possible up to their damage threshold (~ 5 W).[19] This power level is sufficient to pump an OPA. After external pulse picking and compression, 20 W are used to pump a hybrid collinear/noncollinear OPA (Light Conversion Orpheus-F) to generate tunable wavelength pump pulse spanning the UV to the NIR, with pulsewidths from 40 – 120 fs. White light generation (WLG) of a supercontinuum probe pulse is accomplished by focusing a portion (~ 1 W) of unmodulated compressed output directly from the Carbide laser onto a yttrium aluminum garnet (YAG) crystal. For comparison to a conventional transient optical system, we use a 7 W amplified Ti:Sapphire Laser (Coherent Astrella) with a maximum pulse energy of 7 mJ at 1 kHz, center wavelength of 800 nm, and pulse duration of approximately 100 fs.

We employ a standard transient absorption geometry (**Fig. 1**) in which the pump and probe pulses are offset by an angle of less than 10 degrees. Probe pulses are mechanically delayed in time relative to the pump pulse using a delay stage and retroreflector before white light generation. The white light probe is focused onto a sample using a convex mirror to minimize chirp while pump pulses are focused using a long focal length spherical lens. The size of the pump spot is adjusted to be approximately five times larger than the probe spot size. The transmitted pump beam is blocked and the transmitted white light probe is focused onto an optical fiber connected to a miniature spectrometer and fast line-scan camera (Teledyne e2V Octoplus USB) with a pixel height of 200 μm, and a maximum acquisition frequency of 130 kHz. As an additional point of comparison, we use a transmission grating spectrometer (P&P Optica) and a fast InGaAs line camera (Sensors Unlimited 1024-LDH2) with a pixel height of 500 μm, and a maximum acquisition frequency of 92 kHz. Pump and probe intensities are adjusted using a set of neutral density filters. Low repetition rate transient optical measurements are collected using the Ti:Sapphire laser system and a commercial transient absorption spectrometer (Ultrafast Systems Helios). We note that the commercial spectrometer implements a reference detector to correct for pulse-to-pulse fluctuations in the data. Our high-speed system employs a single detector. A statistical analysis comparing the two systems before modulation schemes are applied are included in Supplemental Information. In general, we find that shot-to-shot fluctuations are low compared to time scales that are longer than the laser period. This is the ideal scenario under which maximum benefit can be derived from shot-by-shot detection schemes.

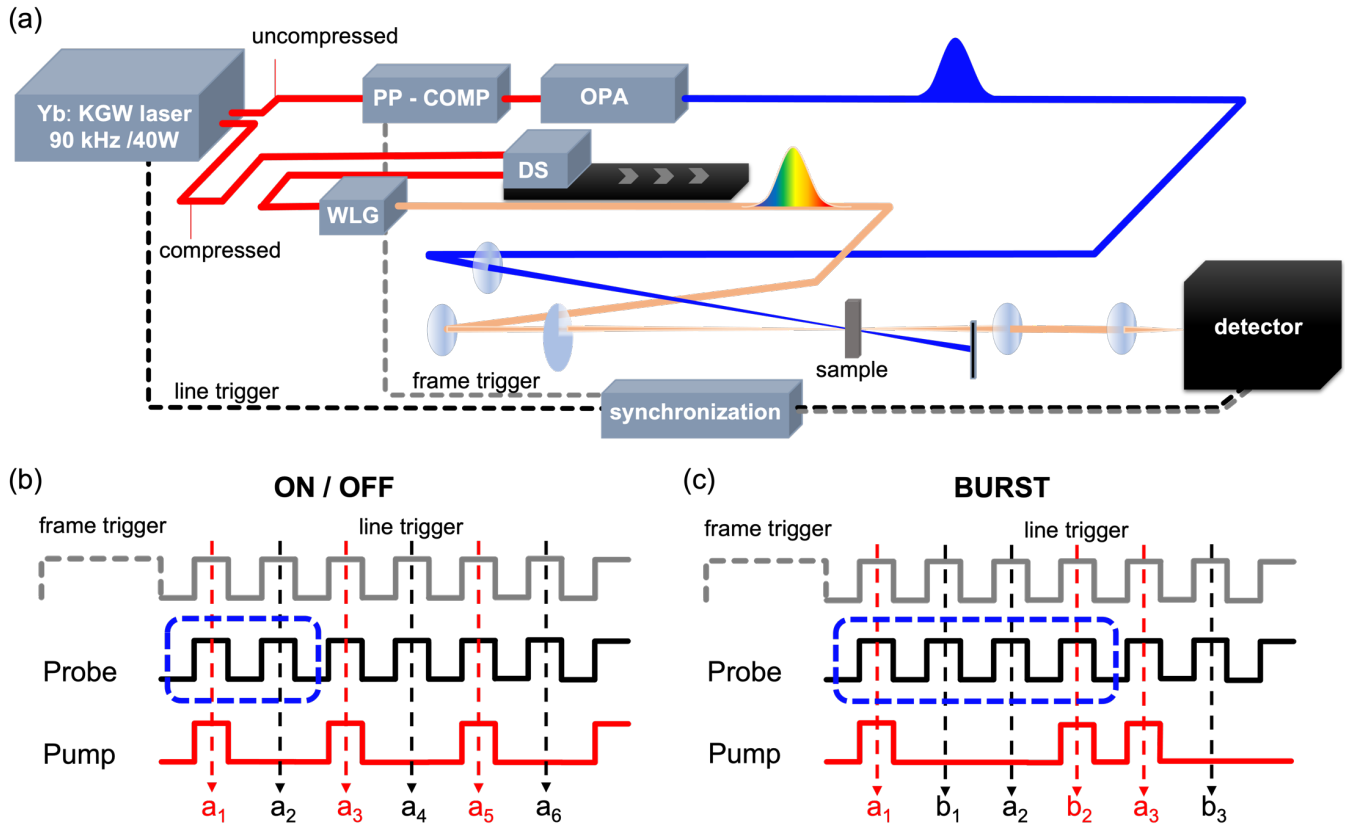

FIG. 1. (a) Schematic of the high repetition rate transient optical measurement. WLG: White light generator; PP: pulse picker; COMP: compressor; DS: delay stage; OPA: optical parametric amplifier. Synchronization of the high-speed detector to the laser pulse train is accomplished using a set of pulse delay generators. A line trigger derived from the laser amplifier synchronizes the shot-to-shot period of the laser to the camera frame rate. A frame trigger drives the pulse picker to set the phase of the measurement and starts data acquisition. (b) Electronic triggering (gray), probe (black), and pump (red) pulse train for conventional "ON/OFF" shot-to-shot modulation and processing in which samples are excited with every alternate probe pulse and data sets are derived from sets of two consecutive probe pulses (blue dotted line). Shot-to-shot acquisition is used to capture the intensity of every pulse ($a_n$). (c) The BURST modified modulation and processing sequence for the electronic triggering (gray), probe pulses (black), and pump pulses (red). Excitation from the pump "bursts" in a sequence of two pump ON and two pump OFF. Shot-by-shot acquisition to generate data sets derived from sets of four consecutive pulses (blue dotted box). This compensates for offsets to the probe intensity affecting sequential pulses that results from mismatched, interleaved ADC signals (labeled as $a_n$ and $b_n$).

**B. Data Acquisition**

To achieve shot-to-shot acquisition, the output of the laser is synchronized to the electronic gate on the camera sensor to set the line rate and compensate for pulse jitter. Furthermore, synchronization between the laser, pulse picker, and camera ensures a consistent pump pulse sequence with a known phase, e.g., the measurement always starts with the "pump-on" condition. The Octoplus camera uses a single input that receives a mixed signal composed of both the "line" trigger, which controls the line rate, and the "frame" trigger that starts data acquisition. The line trigger is taken directly from the sync signal of the laser amplifier and passed through a digital delay board (AeroDiode Tombak) to modify the delay time, amplitude, and pulsewidth. The frame trigger is derived from the same signal that drives our pulse picker. Briefly, a second digital delay board generates a sub-harmonic of the sync signal, which is then routed to two additional delay boards. The third digital delay board provides an additional "burst" signal, permitting arbitrary pulse sequences to drive the pulse picker electronics. The fourth digital delay board is used to select an additional subharmonic to generate the frame trigger for the camera. The product of the two subharmonics is set to achieve the desired acquisition time, typically 0.1 to 4 seconds. In addition, the fourth delay board sets the pulse width for the frame trigger to be twice that of the line trigger so that they are easily distinguished by the camera electronics. The mixed trigger signal (upper trace in **Fig. 1(b)** and **(c)** is obtained by combining the output of the first and fourth delay boards using a resistive power splitter (Mini-Circuits ZFRSC-42-S+).

The Octoplus camera is interfaced directly through a USB port. Data sets are acquired into memory buffers on the camera and streamed directly to a PC for storage and processing. A circular buffer mode ensures that large data sets can be acquired, transferred to the PC, and then requeued without loss. Notably, no specialized data acquisition hardware, e.g. CameraLink, is needed to run the Octoplus camera at full speed. Using the vendor provided software development kit, we built an integrated data acquisition and processing program in C and Python to set parameters for the delay boards, move the mechanical delay stage, retrieve the camera buffers, and display the data. The InGaAs camera is also synchronized to the line trigger and interfaced with a National Instruments frame grabber board.

## C. Data Processing

Conventional shot-to-shot transient optical measurements use an "ON/OFF" sequence, in which the pump pulses are modulated at half the frequency of the probe pulses (**Fig. 1(b)**). A single data point is derived from the signals obtained in two consecutive shots, e.g., $a_n$ and $a_{n+1}$. If the phase is defined so that the data set starts with the pump-on condition, the then transient transmission signal for $N$ laser shots is defined as:

$$\frac{\Delta T}{T} = \frac{2}{N} \sum_{i=0}^{N} \frac{(a_{2n} - a_{2n+1})}{a_{2n+1}}. \quad (1)$$

In typical applications, this shot-to-shot detection scheme represents the lowest noise approach due to the high correlation between subsequent shots.[14,15,17]

However, when the "ON/OFF" approach is applied to data acquired by the Octoplus camera, we observe time-domain fixed pattern noise that is on the same order as typical transient signals (~$10^{-3}$). This effect can be readily seen by comparing two consecutive data sets of 40,000 white light probe pulses. An average of all shots shows that the supercontinuum spectrum is structured and spans the visible and NIR (**Fig. 2(a)**). The maximum intensity near 565 nm has been adjusted to fill 70% of the dynamic range of the camera in a single shot. To characterize the baseline noise, the pump is blocked and a modified version of a transient signal is calculated that is normalized to the full 12 bit depth of the sensor:

$$\Delta S_\theta = \frac{1}{N} \sum_{i=0}^{N} \frac{(a_i - a_{i+1}) e^{i\theta}}{2^{12}}. \quad (2)$$

Here, $e^{i\theta}$ is the phase with $\theta = \pi$ ($\theta = -\pi$) corresponding to a frame that starts with the pump ON (OFF) condition.

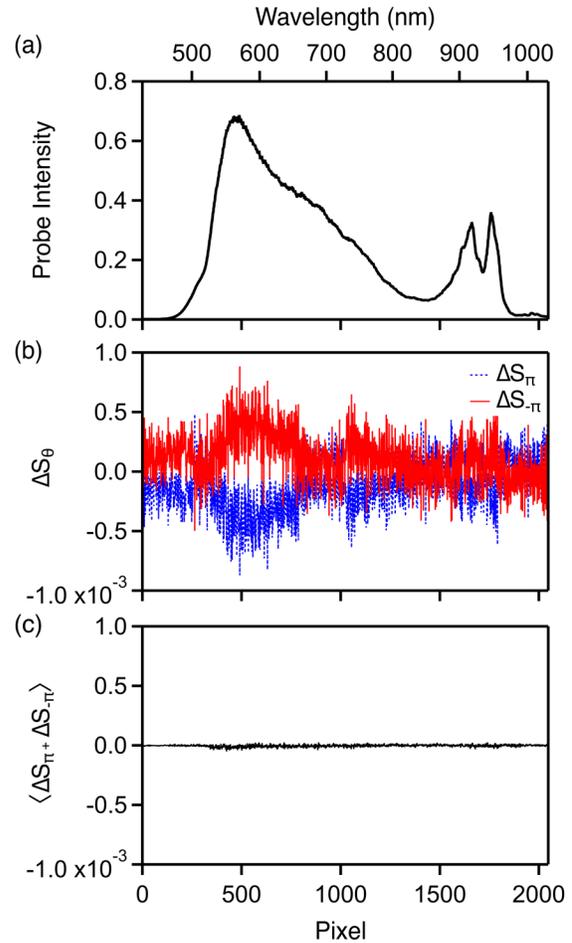

FIG. 2. (a) Average probe spectrum over 40,000 shots recorded with the Octoplus camera. Each shot normalized to the dynamic range of the camera. The maximum intensity of at 565 nm was adjusted to fill approximately 70% of the dynamic range. (b) The differential transmission normalized to the full well capacity of the camera ($\Delta S_\theta$) is determined using shot-to-shot detection and ON/OFF modulation. Two consecutive frames of 40,000 shots each are taken with different phases corresponding to data acquisition starting with pump ON ($\theta = \pi$, blue) versus pump OFF ($\theta = -\pi$, red). A nonzero baseline with a maximum of $10^{-3}$ is observed whose sign switches with the phase of the measurement. (c) A zero baseline with noise below $5 \times 10^{-5}$ is recovered after averaging the two frames ($\langle \Delta S_\pi + \Delta S_{-\pi} \rangle$) in (b).

In the absence of time-domain artifacts, $\Delta S_\theta$ should be centered around zero, decrease with increasing N, and be independent of the phase of the measurement. In contrast, we observe a consistent feature correlated with the maximum of the white light probe intensity, whose sign depends on the phase of the measurement, and whose magnitude is approximately independent of the acquisition time. This pattern is characteristic of "interleaving spurs" that result from camera designs in which high line rates are achieved by interleaving two independent ADCs with a slight mismatch.[20,21] However, taking the average of two consecutive frames with opposite phase $\langle \Delta S_\pi + \Delta S_{-\pi} \rangle$ (**Fig. 2(c)**) results in a flat baseline with noise that is more than an order of magnitude smaller than what is achieved using the "ON/OFF" scheme. This suggests that the detrimental effect of the time-domain fixed pattern noise can be mitigated with a modified data processing approach.

To circumvent issues stemming from interleaving spurs, we implement an alternate "BURST" pump modulation scheme in conjunction with shot-to-shot detection. In BURST mode, the pump is modulated at half the repetition rate of the probe but sequenced such that two consecutive ON shots are followed by two consecutive OFF shots (**Fig. 1(c)**). As each shot is collected and digitized, a transient data point consists of four consecutive probe spectra. The measurement has been phased such that each data point contains an ON shot (collected by ADC $a$), two consecutive OFF shots ($b$ and $a$ respectively), followed by an ON shot ($b$). For $N$ shots, the final transient consists of the average of $N/4$ consecutive data points:

$$\frac{\Delta T}{T} = \frac{4}{N} \sum_{i=0}^{N} \frac{(a_{4i+1} + b_{4i+2}) - (b_{4i+1} + a_{4i+2})}{(b_{4i+1} + a_{4i+2})}. \quad (3)$$

This method effectively removes artifacts related to unmatched ADCs and retains the benefits of the shot-to-shot detection, namely that it groups signals with high temporal correlations. This approach can be adapted to work with the circular buffer design of the Octoplus camera, allowing for arbitrarily long data collection times.

The presence of spurious signals due to time-domain fixed pattern noise is likely to affect a variety of high-speed camera models. For example, we also tested one of the fastest available commercial NIR line cameras (Sensors Unlimited 1024-LDH2) which features an InGaAs sensor with a maximum line rate of 90 kHz. Like the Octoplus CMOS camera, the InGaAs camera also shows interleaving spurs whose effect can be mitigated by changing the pump modulation scheme from ON/OFF to BURST. We tested this camera using the NIR tail of our supercontinuum probe, which contains spectral content extending beyond 1200 nm (**Fig. 3(a)**). Here, instead of normalizing our differential transmission data to the full well capacity, we calculate the full $\Delta T/T$ signal that will exhibit more pronounced noise effects when probe signal intensities are small. We again find a large difference in the effective noise comparing BURST to ON/OFF modulation (**Fig. 3(b)**). For ON/OFF modulation, baseline noise in the $\Delta T/T$ signal extends over a range of approximately 1%, though the noise near the peak of the probe spectrum is an order of magnitude smaller. In contrast, the range of baseline signal using BURST modulation approach is below $10^{-4}$ across the entire spectrum.

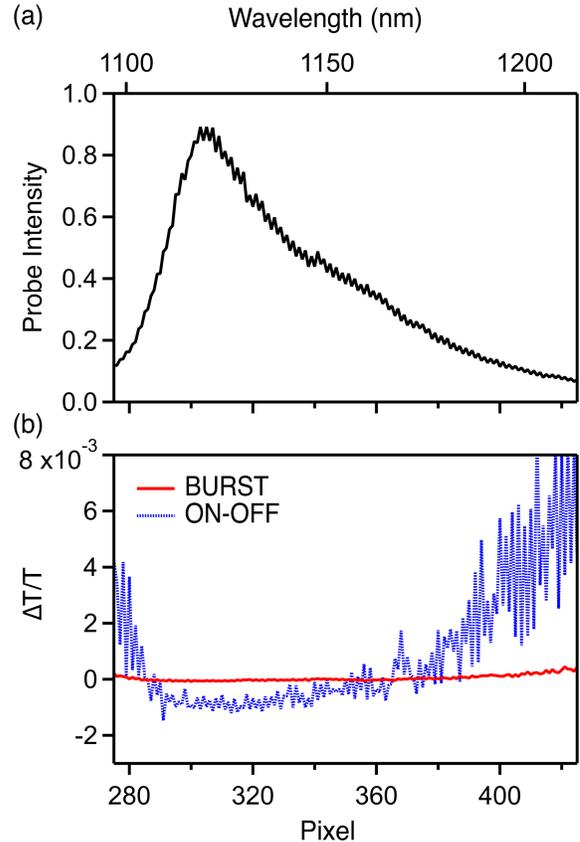

FIG. 3. (a) Probe spectrum averaged over 40,000 shots with the 1024-LDH2 InGaAs line camera using shot-to-shot detection. The peak intensity of a single pulse was adjusted to fill approximately 90% of the full well capacity. The sharp roll off on the blue side results from the spectral filter used to remove the 1030 nm fundamental. (b) The differential transmission signal using the "ON/OFF" scheme shows that noise resulting from interleaving spurs is largely removed using the "BURST" modulation scheme.

## III. PERFORMANCE OF HIGH-SPEED TRANSIENT OPTICAL MEASUREMENTS

### A. Noise Comparisons

Our system exhibits significantly improved acquisition times and lower noise when compared to a commercially available spectrometer designed for low repetition rate operation. To benchmark the performance, we compare our Yb-based system running at 90 kHz in BURST mode to the commercial Helios Spectrometer (Ultrafast Systems) coupled to a 1 kHz Ti:Sapphire system. Shot-to-shot detection is used in both cases and data from the Octoplus camera is binned by a factor of two to match the size of the array detector in the Helios spectrometer (1024 pixels). The systems show comparable noise levels in data sets consisting of the same number of shots (**Fig. 4(a)**). The 1 kHz Helios system takes 4 seconds to acquire 4000 shots and achieves a baseline standard deviation of $\sigma = 8.8 \times 10^{-5}$. The 90 kHz system performs slightly better, acquiring 4000 shots in 0.044 seconds to achieve a baseline standard deviation of $\sigma = 6.3 \times 10^{-5}$. This is consistent with previous reports highlighting the higher stability and correlations of Yb:KGW amplifiers compared to Ti:Sapphire.[17] Considering acquisition time alone, we are able to realize the full 90× speed increase. More importantly, we are now able to fully realize the statistical advantages from collecting a large number of shots when we integrate the high repetition rate system over the same time interval as the low repetition rate system. For example, in four seconds, the 90 kHz system acquires 360,000 shots and achieves a baseline standard deviation of $\sigma = 8.9 \times 10^{-6}$ (**Fig. 4(b)**). This represents a noise reduction by a factor of approximately 9.9× compared to the Helios system, slightly better than what would be expected from statistics alone ($\sqrt{90} \approx 9.5$).

TABLE I. Relative noise contributions to the measurement.

|  | Electronic Noise | Shot Noise | Laser Noise |
|---|---|---|---|
| value | Full Well Capacity $= 2 \times 10^5$ | $N = 1.94 \times 10^5$ | Mean Intensity = 2813 counts |
| error | 50 e- | $\sqrt{N} = 441$ | $\sigma = 45$ |
| relative error | $2.5 \times 10^{-4}$ | $2.3 \times 10^{-3}$ | $1.6 \times 10^{-2}$ |

Similar to previous reports describing noise contributions to transient spectroscopy, we find that laser fluctuations primarily dictate the overall noise level. We follow standard procedures to determine the relative error from electronic, shot, and laser noise.[14] The electronic noise of the camera is determined by the ratio of the absolute error (50 electrons), dictated by the read-out noise, to the full well capacity of each pixel (200,000 electrons). Shot noise is determined by considering that the average quantum efficiency is approximately 72% for this device. Assuming a maximum probe intensity of 70% of the full dynamic range, the number of incident photons per pixel N is 200,000 x 0.7/0.72 = 1.94 x $10^5$. The shot noise is the square root of this value. Knowing the relative errors of the electronic and shot noises, contributions from laser fluctuations can be measured over a finite time. Here, we measure the maximum signal over 40,000 consecutive white light shots and determine the average and standard deviation. We find the relative error to be on the order of 1.6 x $10^{-2}$, resulting primarily from drift. As such, there is significant benefit to be gained from shot-to-shot detection to build data sets from laser shots with the highest correlation.[14,17] Relative errors are summarized in **Table I**.

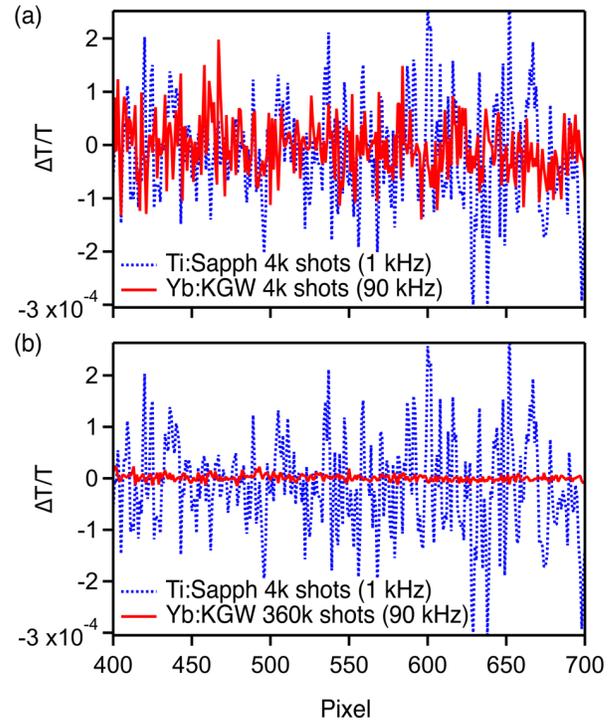

FIG. 4. (a) Baseline noise comparison for the same number of shots between a 1 kHz Ti:Sapphire based commercial transient absorption spectrometer and the 90 kHz Yb:KGW based system. The two exhibit nearly equivalent noise levels over 4000 shots. (b) Baseline noise comparison over a 4 second acquisition time between a 1 kHz Ti:Sapphire based commercial transient absorption spectrometer (4000 shots) and the 90 kHz Yb:KGW system (360,000 shots). The standard deviation for the high repetition rate setup is nearly an order of magnitude smaller than the Ti:Sapphire setup.

## B. Differential transmission of graphene at low fluences

Fluence-dependent transient measurements of graphene are an effective tool for demonstrating the sensitivity of the high repetition rate transient system. Graphene is 2D semimetal with an approximately linear dispersion for low energy charge carriers and a vanishing density of states near the Dirac point.[22] As a result, low energy electrons in graphene have zero effective mass and can propagate relatively long distances without scattering. Consequently the photophysics at room temperature is of broad interest and relevant to a range of optoelectronic devices.[23–26]

In graphene, the magnitude and sign of the differential transmission signal strongly depends on the incident laser fluence and the probe wavelength. The carrier dynamics can be reasonably approximated using a model that partitions the electrons and phonons into independent sub-systems that equilibrate through the emission of optical phonons. As a result, the excited state dynamics are governed by a few characteristic timescales corresponding to (i) electron thermalization in tens of femtoseconds,[27,28] (ii) carrier lattice equilibration via electron-phonon coupling,[29] which can include substrate interactions[30,31] and (iii) lattice cooling via anharmonic decay of optical phonons in a few picoseconds.[32,28] The observed differential transient signal is a sum of all these processes. At early times (< 1 ps), the net transient signal reflects contributions with opposing signs: interband state blocking from hot electrons contribute a positive $\Delta T/T$ signal and phonon assisted intraband absorption contributes a negative $\Delta T/T$ signal. As the relative amplitude and decay time of the positive interband contribution depends on the initial electron temperature, the overall transient signal will be a function of the laser fluence.[22] These effects have been described theoretically[33] and observed using degenerate pump-probe measurements.[29] For example, at high fluences, the net signal is positive for all times. As the fluence is lowered, a cross-over regime is observed such that a net negative $\Delta T/T$ emerges at finite delay times. Here, we use this behavior to assess the sensitivity of our transient system and demonstrate a novel data set resulting from achieving high SNR in ultrafast measurements.

We find that the low-fluence cross-over regime in graphene can be readily observed using broadband differential transmission measurements. Here our sample consists of large area monolayer graphene grown by chemical vapor deposition on glass (Graphenea). The sample was excited with a 450 nm pump with a spot size of 415 μm. To facilitate comparison with previous studies,[29] we initially focus on the kinetics for probe wavelengths near 850 nm (**Fig. 5**). At high fluences above 10 μJ/cm² per pulse, we observe a positive $\Delta T/T$ signal for all times. This results from the initially high electron temperature that results in Pauli blocking of interband transitions. Below the critical fluence, we observe a crossover behavior in the normalized kinetics (**Fig. 5(a)**), in which a negative $\Delta T/T$ signal emerges after a few hundred femtoseconds, consistent with rapid hot carrier thermalization followed by the onset of intraband absorption. Of particular note is the measurement taken at a fluence of 0.65 μJ/cm², where the maximum observed signal at 850 nm is only 6 x 10⁻⁶ (**Fig. 5(b)**). At this carrier density, the magnitude of the negative signal $\Delta T/T$ is as large as the magnitude of the early time positive $\Delta T/T$ signal. We emphasize that these low $\Delta T/T$ signal levels achieved using broadband multichannel detection are among the lowest reported to date on monolayer graphene using any pump-probe system. For example, our sensitivity rivals two-color measurements taken with MHz oscillators[31] and is at least an order of magnitude higher than that of the best reported two-color measurements taken with amplified systems.[29,34] We note that our measurements will also include noise contributions from fluctuations in the pump intensity with time that are not accounted for in the above analysis. However, we find that long term drift is negligible over dozens of sweeps and that shot-to-shot pump fluctuations contribute minimal additional noise on average.

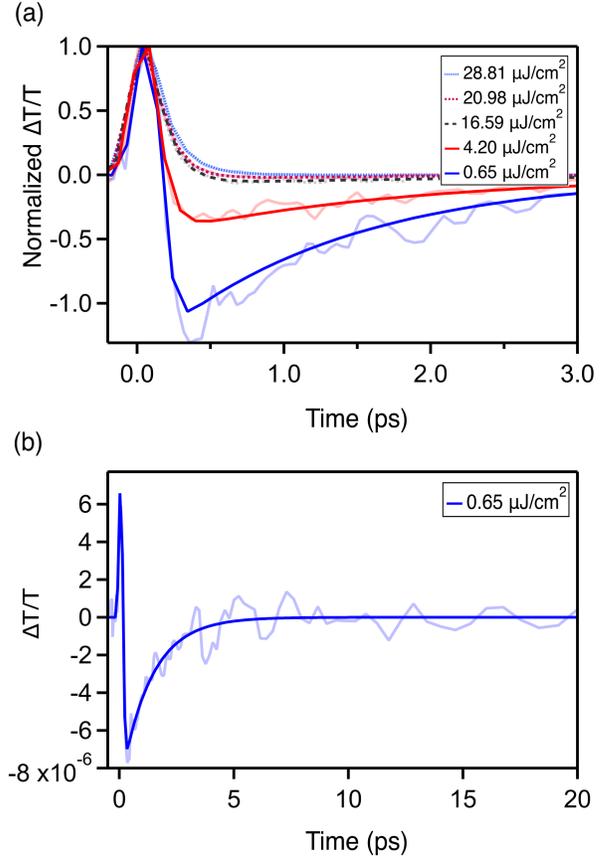

FIG. 5. (a) Normalized differential transmission kinetics from monolayer graphene pumping at 450 nm and probing at 850 nm for six fluences: 28.81 μJ/cm², 20.98 μJ/cm², 16.59 μJ/cm², 4.20 μJ/cm², and 0.65 μJ/cm². The net sign of the signal for delay times between 0.1 – 1 ps changes from positive to negative as the fluence is lowered. (b) The maximum signal for the lowest fluence excitation is only 6 x 10⁻⁶.

Furthermore, our measurement exhibits the additional benefits provided by using multichannel detection. For example, the spectrally resolved probe allows us to clearly see variations in the intensity and sign of the signal as a function of probe wavelength. As thermalized electrons in graphene follow Fermi-Dirac statistics, the intensity and lifetime of the positive interband signal should increase as the probe photon energy decreases. The raw broadband differential transmission data over the range of 600 – 900 nm (**Fig. 6(a)**) clearly confirm this scaling of the magnitude of the positive $\Delta T/T$ signal, which ranges from ~ 1 x 10⁻⁶ at 600 nm to 13 x 10⁻⁶ at 920 nm. Additionally, the broadband data set is amenable to multidimensional global analysis methods that permit high fit convergence and noise reduction. Here, a simple three state sequential model is applied to our data set to capture its essential characteristics. Using singular value decomposition, we extract three linearly independent species and three rate constants, which can then be used to reconstruct the full transient data set with significantly reduced noise (**Fig. 6(b)**). The residual difference between the raw data and the fit (**Fig. 6(c)**) shows that three time constants are sufficient to capture the essential dynamics in our system. The first time constant, corresponding to electron thermalization, is fixed to the instrument response of the system. The second time constant represents the lifetime of the optical phonon and is determined to be 1 ps. In agreement with previous results, the optical phonon lifetime is found to be independent of pump fluence. The final component extends for several picoseconds and corresponds to lattice cooling.

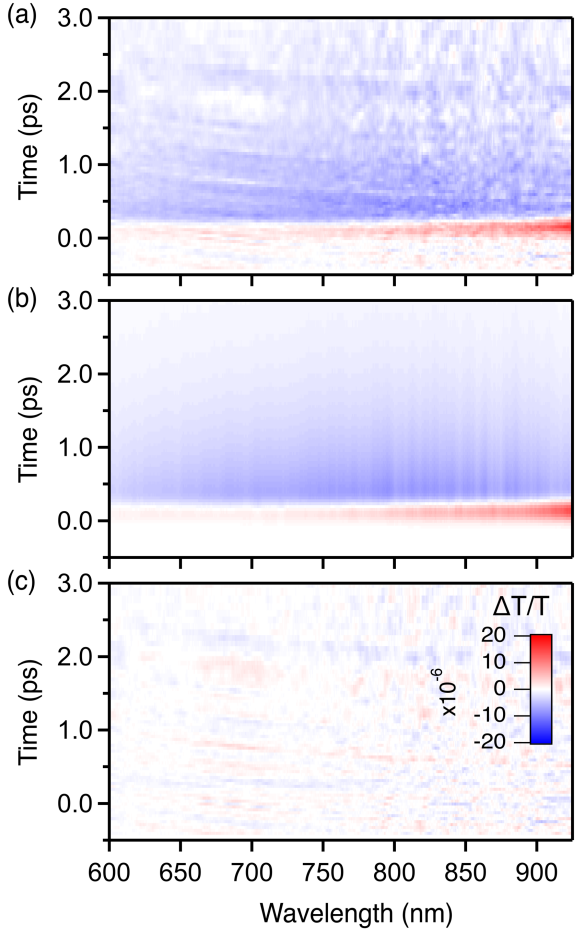

FIG. 6. (a) An image representation of the differential transmission for a pump fluence of 0.65 μJ/cm$^2$ as a function of probe wavelength and time shows that the magnitude of the state blocking contribution increases as the probe wavelength increases. (b) The data set can be fit using a sequential global analysis model to extract three characteristic time constants and filter out additional noise in the measurement. (c) The differential between the raw data and fit shows that the fit successfully reproduces the main features of the data.

## IV. CONCLUSION

We show that differential transmission signals as small as 1 x 10$^{-6}$ can be readily measured in a simplified broadband transient system using electro-optic modulation to achieve shot-to-shot detection. This can be accomplished using an inexpensive camera with direct USB data readout, high acquisition rates, large pixel heights, and arbitrarily long acquisition times. However, the dual ADC design of these cameras requires a modified modulation and data processing approach to avoid spurious time-dependent fixed pattern noise signals. We have demonstrated that this tool is uniquely effective for investigating the low fluence dynamics regime in low-dimensional materials. Using monolayer graphene, we show that we can preferentially detect the intraband absorption signal over a wide range of probe wavelengths and use multidimensional data processing approaches to obtain low noise and high fit convergence.

## V. ACKNOWLEDGMENTS

This work was supported by the National Science Foundation under grant DMR-2004683. Support for P.B. was provided by the Lindemann Trust Fellowship. Support for A.Z. was provided by a PSC-CUNY Award, jointly funded by The Professional Staff Congress and The City University of New York. Support partially provided by the National Science Foundation NSF-CREST Center for Interface Design and Engineered Assembly of Low Dimensional Systems (IDEALS), NSF grant number NSF-CREST HRD-1547830. This research used resources of the Center for Functional Nanomaterials, which is a US DOE Office of Science Facility, at Brookhaven National Laboratory under Contract No. DE-SC0012704.

We gratefully acknowledge assistance from Dan Bloom (Teledyne) with the hardware and software interface of the Octoplus camera. The external pulse picker and compressor was designed by Light Conversion. We especially thank Šarūnas Straigis for help with pulse picker trigger inputs. We additionally recognize contributions from Chris Middleton (PhaseTech Spectroscopy) in identifying the origin of fixed pattern noise in these cameras.

## VI. DATA AVAILABILITY

The data that support the findings of this study are available from the corresponding author upon reasonable request.